\documentclass[prl,twocolumn,superscriptaddress,nofootinbib]{revtex4-1}
\usepackage{amsmath,amssymb,bm}
\usepackage[varg]{txfonts}
\usepackage{mathrsfs} 

\newcommand{\ket}[1]{|{#1} \rangle }

\usepackage{amsmath}	
\usepackage{color}
\usepackage[pdftex]{graphicx}
\usepackage{hyperref}
\begin{document}
	\title{Transient Joule- and (ac) Josephson-like photon emission in one- and two- nucleon tunneling processes between superfluid nuclei: blackbody and coherent spectral functions}
\author{R. A. Broglia}
	\affiliation{The Niels Bohr Institute, University of Copenhagen, DK-2100 Copenhagen, Blegdamsvej 17, Denmark
		}
\affiliation{Dipartimento di Fisica, Universit\`a degli Studi di Milano, Via Celoria 16, I-20133 Milano, Italy}
    \author{F. Barranco}
	\affiliation{Departamento de F\'isica Aplicada III, Escuela Superior de Ingenieros, Universidad de Sevilla, Camino de los Descubrimientos, Sevilla, Spain}
    	\author{G. Potel}
	\affiliation{Lawrence Livermore National Laboratory, Livermore, California 94550, USA}	
	\author{E. Vigezzi}
	\affiliation{INFN Sezione di Milano, Via Celoria 16, I-20133 Milano, Italy}
	\date{\today}
	\begin{abstract}
      Effective charged neutrons involved in one- and two- nucleon tunneling processes in heavy ion collisions between superfluid nuclei are expected to emit photons. Although the centroid, width and integrated energy area characterizing the associated $\gamma$-strength functions are rather similar, the corresponding line shapes reflect the thermal equilibrated-like character of the quasiparticle transfer (1$n$-channel, blackbody spectral functional dependence), and the quantal coherent character of the Cooper pair transfer ($2n$-channel, Gaussian functional dependence) respectively. The predicted angular distributions, polarizations and analyzing power provide further insight into the profoundly different physics to be found at the basis of what can be considered a transient Joule-like and a (ac) Josephson-like nuclear processes.
	\end{abstract}    
	\maketitle
	\section{Introduction}
    A potential (voltage) difference, whether static or time dependent, between two points in a metal creates an electric field that accelerates electrons, giving them kinetic energy.
    
\emph{Room temperature--}
When the charged particles interact with the lattice phonons (electron-phonon coupling), energy and momentum is transferred from the electrons to the lattice by the creation of further phonons, phenomenon closely connected with resistivity. Oscillations of electrons (also of plasmons) and of ions in thermal equilibrium are at the basis of the emission of electromagnetic radiation which, for example, an incandescent light bulb emits.  Radiation also known as blackbody radiation. It includes part of the ultraviolet (UV) and all of the visible and infrared (IR) spectrum.

\emph{Low temperatures--} At very low temperatures, as that associated with liquid helium (boiling temperature 4.23 K that is \mbox{-268.8$^\circ$C}, close to absolute zero temperature), current in many metals, in particular in bad conductors like lead, displaying a rather strong electron-phonon coupling, is carried not by single electrons, but by weakly bound, very extended, strongly overlapping pairs of electrons, known as Cooper pairs \cite{Cooper:56}. The role of phonons, mainly of quantal zero point fluctuation origin, is now that to act as designed glue of Cooper pairs by being exchanged between the corresponding electron partners moving in time reversal states, and not as scattering centers\footnote{Within this context it is of note that, the difference with the finite temperature phenomena is a fundamental one, namely the invasion of the large scale world by quantum mechanics. The two salient pictures of the low temperature world, the vanishing entropy and the dominance of the zero point energy are its direct consequences \cite{Mendelsson:77}.}. Thus conduction of current without resistance, namely superconductivity \cite{Bardeen:57a,Bardeen:57b}. In a superconductor there is no resistance because all Cooper pairs carry the same phase $\phi$, being  collectively in the same state, the coherent $|BCS(\phi)\rangle$ state \cite{Schrieffer:64}. Once one gives a momentum to the center of mass of them and starts a supercurrent, to get one Cooper pair or one electron partner --that is, break a Cooper pair-- away from what all others are doing is very hard. It implies an exchange of energy of about twice the pairing gap\footnote{In other words, an energy of the order of $T_c$, where $T_c$ is the critical temperature at which the phase transition between the normal (N) and superconducting (S) phases takes place. In the case of lead, $T_c\approx7.19$ K (0.62 meV).} (e.g. $\Delta=1.4$ meV for lead), phenomenon known as Off-Diagonal Long-Range Order (ODLRO) \cite{Penrose:51,Penrose:56,Anderson:96}.

    % If one turns a magnetic field on a superconductor there would be a transient electric field which would generate a surface current opposing the flux, as also happens in a metal in the normal phase, in which case the induced current will die away fast (Joule effect). In a superconductor this mechanism is not operative as a supercurrent, once started will keep on going forever (persistent currents), the applied magnetic field being permanently excluded \cite{Meissner:33}.

    Let us now consider  a circuit made  out of two pieces of metal like lead, oxided across their contact surface (weak link; see e.g. Figs 1.1 and 1.5 ref. \cite{Barone:82}), closed by a conducting cable in which a (dc) battery has been inserted. Immerse the circuit in an appropriate fashion in a container filled with liquid helium (see Fig. 3.9 of ref \cite{Barone:82}). Approach a compass to the system. The needle, which before immersion had reacted to the current induced by the potential difference $V$ created by the (dc) battery, with a shift of the needle from the unperturbed South (S)$\to$North (N) (S$\to$N) position to a new one, now indicates the S$\to$N direction, implying that the direct current induced at room temperature by the battery is not circulating anymore.  If one had access to something similar to a (non standard)``radio'' receiver switched on the\footnote{Electromagnetic high frequency/far infrared.} EHF/FIR region of the electromagnetic spectrum (around the THz region), and tuned it very delicately in search for emitters, one will find a rather weak ``station''. It ``broadcasts'' in a single microwave frequency, implying that along the circuit, an alternating current is circulating --so called (ac) Josephson effect \cite{Josephson:62,Anderson:64b}-- and consequently, the weak link is emitting photons of that frequency ($\nu_J=2e\times V/h$, see e.g. ref. \cite{Lindelof:81}).
    \section{(ac) Josephson- and Joule-like nuclear photon radiation}
    A systematic study of one- and two-nucleon transfer reactions in heavy ion collisions between superfluid nuclei, namely
    \begin{align}\label{eq:1}
      		^{116}\text{Sn}+^{60}\text{Ni}\to\left\{\begin{array}{c}
		^{115}\text{Sn}+^{61}\text{Ni}\quad (Q_{1n}= -1.74\text{ MeV}),\text{(a)} \\ [10pt]
		^{114}\text{Sn}+^{62}\text{Ni}\quad (Q_{2n}= 1.307\text{ MeV}),\,\text{(b)}
		\end{array} \right.
    \end{align}
	was reported in refs \cite{Montanari:14,Montanari:16}. These reactions were carried out for twelve bombarding energies, from values above that of the Coulomb barrier to well below it. At these low energies, target and projectile give rise, around the distance of closest approach where their pairing --abnormal-- densities overlap weakly, to a Josephson-like junction  of transient character, as the collision --weak contact-- time is of the order of $\tau_{coll}\approx0.5\times10^{-21}$s \cite{Potel:21,Broglia:21} (see also \cite{Magierski:21}).

    From the analysis of the  data associated with the reaction given in Eq. (\ref{eq:1}) (b), it was estimated \cite{Potel:21} that the mean square radius (correlation length) of the transferred Cooper pair in the {\it ground state} (gs) $\to$ {\it ground state} (gs) transfer process, is $\xi\approx13.5$ fm (see also Fig. 4 of ref \cite{Broglia:21}). A quantity related to the largest relative distance of closest approach --barrier width $d$, i.e. distance between the radius of the nuclear densities ($d\approx\xi-R_0(^{116}\text{Sn})-R_0(^{60}\text{Ni})\approx2.24$ fm) -- for which the absolute two-nucleon tunneling cross section is of the order of the single-particle one (see Fig. 2 \cite{Broglia:21}). 
\begin{figure}
	\centerline{\includegraphics*[width=9cm,angle=0]{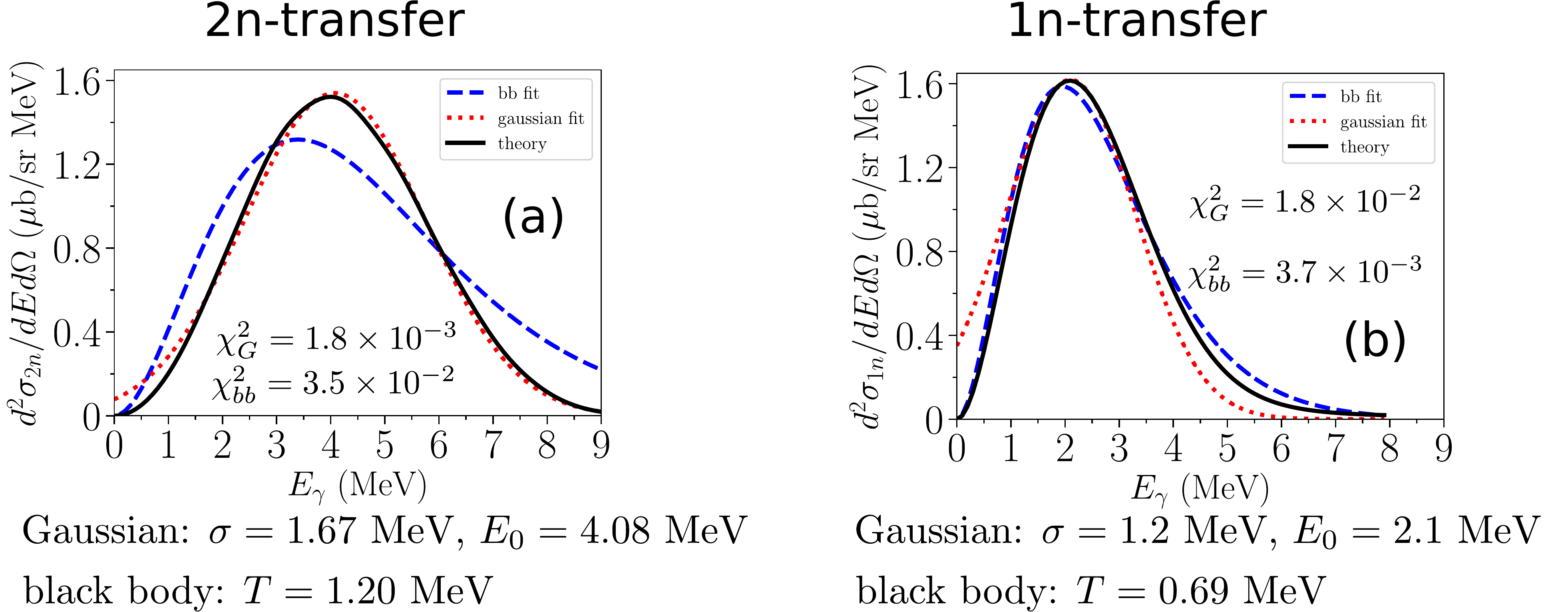}}
	\caption{$\gamma$-strength function for 2$n$- and 1$n$-tunneling and associated blackbody ($\chi^2_{bb}$) and Gaussian ($\chi^2_{G}$) fits. ({\bf{a}}) (continuous black line) $\gamma$-emission (gs)$\to$(gs) two-neutron tunneling (Eq. \ref{eq:1} (b)) absolute double differential cross section for $E_{cm}=154.26$ MeV and $\theta_{cm}=140^\circ$ as a function of the emitted $\gamma$-ray energy $E_\gamma$, calculated with Eq. (\ref{eq:306}); ({\bf{b}}) (continuous black line) $\gamma$-emission in one-neutron tunneling (Eq. (\ref{eq:1} (a))) absolute double differential cross section at the same kinematical conditions as (a), calculated with an expression similar to (\ref{eq:306}) for each of the incoherent quasiparticle contributions (eleven, four being the most important ones), taking properly into account the angular momentum coupling coefficients associated with the  quasiparticle contributions of total angular momentum $j$. Also displayed are the Gaussian (red dotted curve) and blackbody (blue dashed curve) fits, and associated chi-squared values and parameters.}\label{fig1}
\end{figure}
    Furthermore, emission of $\gamma$-rays of (nuclear N ) Josephson-like frequency $\nu^N_J=Q_{2n}/h$ distributed over an energy range of few MeV ($\approx \sqrt{3}\hbar/\tau_{coll}$ \cite{Potel:21}\footnote{Such a radiation implies a momentum exchange $h\nu/c$ \cite{Stachel:87} and thus an associated beam dispersion. In keeping with the fact that the momentum of relative motion of the reaction (\ref{eq:1} (b)) is, in the entrance channel, $k_i\approx1.76$ fm$^{-1}$ for the kinematic conditions selected ($E_{cm}=154.26$ MeV), and that the energy $E_\gamma$ of the most probable photon associated with Cooper-pair tunneling is 4 MeV, the recoil $\gamma$ momentum is $k_\gamma\approx0.02$ fm$^{-1}$ and thus negligible with respect to $k_i$. }) was predicted (Fig. \ref{fig1} (a)). Similar $\gamma$-strength functions are predicted for the one-nucleon (1n) tunneling process (Fig. \ref{fig1} (b)).
    In what follows, it will be argued that such similarity is only apparent. Actually, it constitutes the first and, likely, most important piece of evidence testifying to the fact that the reaction Eq. (\ref{eq:1} (b)) reflects a quantal process between two coherent ($|BCS\rangle$) states describing the (gs) of the initial and final systems, while the reaction Eq. (\ref{eq:1} (a)) populating a number of quasiparticle states with $E_{qp}\lesssim 2.4$ MeV, leads to a $\gamma$-ray distribution reflecting a statistical  process resulting from the incoherent summed contribution of the individual quasiparticle channels. A gas of virtual photons\footnote{The emitted photons can, during the collision time, travel a distance $\tau_{coll}\times c\approx10^2$fm.} related, in the first case Eq. (\ref{eq:1} (b)), with quantal coherent states built out of pairs of nucleons moving in time reversal states all with the same (gauge) phase, and of incoherent quasiparticle states of arbitrary phase in the second case (Eq. (\ref{eq:1} (a))). While it is possible to accurately fit ($\chi^2_{bb}=3.7\times10^{-3}$) the line shape of the $\gamma$-strength function associated with the one-neutron (1$n$) tunneling process making use of the blackbody spectral radiance function
    \begin{align}
      \label{eq:2}
      B_{\nu_\gamma}(h\nu_\gamma,T)=\frac{2h\nu_\gamma^3}{c^2}\frac{1}{e^{\frac{h\nu_\gamma}{kT}}-1},
    \end{align}
    (see Fig. \ref{fig1} (b)), this is not so in the case of the two-neutron ($2n$) $\gamma$-strength function (Fig. \ref{fig1} (a)).

    On the other hand, this ($2n$) strength function allows for an accurate fit ($\chi^2_G=1.8\times10^{-3}$) in terms of a Gaussian shape
    \begin{align}
      \label{eq:3}
      f(E_\gamma)=\frac{1}{\sigma\sqrt{2\pi}}\exp\left(\frac{(E_\gamma-E_{0})^2}{2\sigma^2}\right),
    \end{align}
    --fingerprint of coherent photon emission \cite{Glauber:07}-- something that eludes the ($1n$) $\gamma$-strength function, as seen from Fig. \ref{fig1} (b).

    Within the context of Eq. (\ref{eq:2}) one can ask, how does one make a heat bath in a nucleus? Although it is not a thermal bath in the classical sense, when the $1n$-tunneling ``current'' emits $\gamma$-rays, its energy distribution is determined by the (inclusive) level density of the daughter system (20 MeV$^{-1}$, that is, 40 levels distributed over 2 MeV, with a spectroscopic weighted centroid $\approx1$ MeV \cite{Lee:09}). While this level density is not that of a macroscopic system, neither that of a highly excited nucleus \cite{Bortignon:98}, it suffices to accurately mimic the blackbody electromagnetic emission line shape.    

    The experimental setup used to study the reactions given in Eq. (\ref{eq:1}) (2 MeV beam energy resolution) implies inclusive reaction conditions for both processes (Eq. \ref{eq:1}(a)) and (Eq. \ref{eq:1}(b)), each of which has also the full phase space available for $\gamma$ emission. However, it is only (Eq. \ref{eq:1} (a)) which proceeds through the (inclusive), incoherent channel (Giaever-like regime \cite{Giaver:73}; see also Fig. 1 of \cite{Broglia:21}). In fact, the process (Eq. \ref{eq:1} (b)) can be accurately accounted for in terms of a single coherent quantum transition between the ground states of initial and final nuclei. As a consequence, the Gaussian line shape of the associated $\gamma$-strength function\footnote{It is of note that the extreme sensitivity of the process to the nuclear density of levels is reflected in the fact that in moving from a single coherent quantal transition to the incoherent sum of four quasiparticle quantal transitions, the associated $\gamma$-strength function line shape changes from a Gaussian to a blackbody radiation functional dependence. Within this scenario, one can only speculate on its consequences in the quest of the statistical aspects of nuclei at modest excitation energies like, for example, the region of the pygmy dipole resonance probed with two-nucleon transfer processes (see e.g. \cite{Broglia:19c} and refs. therein).}.

    The parallel of the above processes with the electromagnetic radiation emitted by a glowing light bulb over a wide range of frequencies, and the radiation emitted by a laser which is essentially monochromatic, is apparent. The reason for the similar FWHM of the $(1n)$- and $(2n)$-$\gamma$-strength functions is in keeping with the transient character of the interaction between projectile and target ($\tau_{coll}\approx0.5\times10^{-21}$s), and the recoil effect --change of trajectory of relative motion-- associated with the tunneling processes, which blurs the density of levels of the final system, although not its line shape. It is of note that the energy  of the emitted $\gamma$-rays is taken  from the relative motion when the $Q$-value of the reaction is negative as is the case of Eq. (\ref{eq:1} (a)), or when $E_\gamma>Q_{2n}$ as in Eq. (\ref{eq:1} (b)) ($\delta$-function, Eq. (\ref{eq:306})).

    To further support of the above parlance, one finds the close numerical resemblance of the energy integrated area of the $(2n)$- and $(1n)$-$\gamma$-strength functions, which testifies to the fact that the coherent $\ket{BCS}$ initial and final states of the $(2n)$-tunneling process are built out of the same single-particle phase space associated with the $(1n)$-quasiparticle incoherent channel.

    Within this context it seems appropriate to mention, of the parallel existing between the fact that already a BCS condensate of few (4--6) nuclear Cooper pairs displays many of the properties of superfluids, with the fact that the incoherent contribution of already four quasiparticle states, excited in one-nucleon transfer reactions, emits an ensemble of photons     which very much resemble blackbody radiation, controlled by an effective temperature.

    In the explanation given by Einstein \cite{Einstein:07} of the low temperature specific heat, he used a model of a crystal in which temperature is given by the energy with which  the lattice vibrates, represented by harmonic oscillators, all with the same characteristic frequency $\nu_0$. It was noted that such a model bears a close resemblance to the mathematical framework which Planck had used in the derivation of the radiation formula. There too, deviations had occurred from the law of equipartition, especially at low frequencies (temperatures). Within this context the relation $T_{char}=h\nu_0$, where Boltzmann's constant has been set equal to 1, defines the characteristic temperature at which the thermal energy per degree of freedom becomes equal to the value of one energy quantum. Below this temperature the probability of the degrees of freedom receiving one quantum decreases rapidly, the opposite being true above $T_{char}$. According to the above parlance and in connection with the one-quasiparticle transfer process given in Eq. (\ref{eq:1}) (a) one    is, arguably, dealing with a situation which parallels that Einstein treated in connection with low-temperature specific heats.

    The interplay between structure and reactions found in the process expressed in Eq. (\ref{eq:1} (a)) implies that, in principle, there are different ways in which one can estimate the value of $T_{char}$: a) making use of the equipartition principle for a gas of point particles\footnote{This in keeping with the fact that the reaction process is controlled, at bombarding energies well below the Coulomb barrier, essentially by the real part of the optical potential and the Coulomb interaction. That is, mainly by a central potential, with no spin-orbit contribution.}, that is, $(3/2)T_{char}=E_{cm}/A$ leading to $T_{char}\approx (2/3)154.26$ MeV/176 $\approx 0.58$ MeV; b) from the transfer cross section ($\sigma_j$) average quasiparticle energy $T_{char}=\langle E\rangle=\sum_{j}(\sigma_j E_j(qp))/\sum_j\sigma_j$, $E_j(qp)$ being the $^{61}$Ni quasiparticle energies,  while the denominator, is taken from the experimental data (7.25 mb/sr), and the single values $\sigma_j$ from theoretical estimates, renormalized by the ratio $(\sum_j \sigma_j)_{exp}/(\sum_j \sigma_j)_{th}\approx 7.25/6.03\approx 1.2$ (Table  1 of \cite{Potel:21}, $D_0=13.49$ fm), in which case $T_{char}\approx$ 0.67 MeV; c) making use of the spectroscopic factor averaged quasiparticle energy $T_{char}=\langle E\rangle=\sum_{j}(S_j E_j(qp))/\sum_j S_j\approx 1.01$ MeV,  the $^{61}$Ni quasiparticle energies fulfilling in this case the condition to be  $\lesssim 2$ MeV, value which acts as a cutoff  (role played in b) by the decreasing values of $\sigma_j(Q_{1n}-E_j(qp))$ as a function of $E_j$ ), $S_j$  being the corresponding spectroscopic factors \cite{Lee:09}; d) from the fitting of the quantum mechanical calculated $\gamma$-strength function displayed in Fig. \ref{fig1} (b), in which case $T_{char}\approx 0.69$ MeV. It is of note  that the resulting values of $T_{char}$ are  similar, leading to an average value of $0.74\pm0.16$ MeV.

Concerning the line shape associated with the coherent infrared and optical oscillator stimulated emission (maser, laser) we refer to \cite{Schawlow:58}  (see also \cite{Townes:65,Prokhorov:65,Basov:65}). Doppler broadening and spontaneous emission related to zero point fluctuations result in different full width at half maximum power emission and line profiles.   
    
    In what follows we discuss a second evidence of the parallel between ($2n$)-(coherence, Josephson-like) and ($1n$)-(blackbody, Joule-like), in terms of the angular distribution and of the polarization of the emitted $\gamma$-rays in the processes given in  Eq. (\ref{fig1} (b)) and Eq. (\ref{fig1} (a)) respectively.

\section{Angular distribution and polarization}	
    Let us first consider the two-nucleon tunneling processes $a+A\to b+B$ in which two neutrons are transferred from the even-even nucleus $a(\equiv b+2)$ to the even-even nucleus $B(\equiv A+2)$, and  circular polarized photons of energy $E_\gamma$ are emitted in the direction $\hat k_\gamma$. The $T$-matrix can be written as,
    \begin{align}
      \label{eq:300}
\mathcal T^q(\mathbf k_\gamma,\mathbf k_f)=\sum_{m_\gamma} \mathcal D_{m_\gamma q}^1(R_\gamma)\,T_{m_\gamma}(\mathbf k_f) ,        
    \end{align}
    where $q=\pm1$ is the photon polarization, $\mathcal D_{m_\gamma q}^1(R_\gamma)$ are the Wigner matrices describing the rotation from the quantization axis to the direction $\hat {\mathbf k}_\gamma$, while   $T_{m_\gamma}(\mathbf k_f)$   describes the $\gamma$-emission of the successive transfer of neutrons, overwhelming contribution to the associated absolute differential cross section, is written as
    \begin{widetext}
\begin{align}\label {eq:301}
\nonumber T_{m_\gamma}(\mathbf k_f)&=\sum_{j_i,j_f}B_{j_i}B_{j_f}\int \chi_f^*(\mathbf r_{Bb};\mathbf k_f)\left[\phi_{j_f}(\mathbf r_{A_1})\phi_{j_f}(\mathbf r_{A_2})\right]^{0*}_0  D_{m_\gamma} \left[\phi_{j_f}(\mathbf r_{A_2})\phi_{j_f}(\mathbf r_{b_1})\right]^{K}_M\,U^{(A)}(r_{b_1})\, d\mathbf r_{Cc}\, d\mathbf r_{b_1}\, d\mathbf r_{A_2}\\
&\times\int G(\mathbf r_{Cc},\mathbf r'_{Cc})\left[\phi_{j_f}(\mathbf r'_{A_2})\phi_{j_i}(\mathbf r'_{b_1})\right]^{K*}_M U^{(A)}(r'_{c_2})\left[\phi_{j_i}(\mathbf r'_{b_2})\phi_{j_i}(\mathbf r'_{b_1})\right]^{0}_0\chi_i(\mathbf r'_{Aa})\, d\mathbf r'_{cC}\, d\mathbf r'_{b_1}\, d\mathbf r'_{A_2}.
\end{align}
\end{widetext}
The dipole operator is defined as,
\begin{align}
  \label{eq:302}
   D_{m_\gamma}=e_{eff}\sqrt{\frac{4\pi}{3}}\left(r_{O1}Y^1_{m_\gamma}(\hat r_{O1})+r'_{O2}Y^1_{m_\gamma}(\hat r'_{O2})\right),
\end{align}
While the coefficients $B_{j_i},B_{j_f}$ are the BCS coherence  factors ($U_\nu V_\nu$) describing the ground states of the nuclei $a$ and $B$, respectively.  The effective neutron charge is $e_{eff}=-e\,\frac{(Z_A+Z_b)}{A_A+A_b}$. 
For a numerical evaluation of Eq. (\ref{eq:301}),  $T_{m_\gamma}(\mathbf k_f)$ is expanded in partial waves.

Because nucleons carry effective charges, transfer tunneling processes like (Eq. \ref{eq:1} (b)) can be viewed  as a current of carriers of charge $2\times e_{eff}=-e\times 0.89$. Thus, the standard expression of the $T$-matrix should contain the dipole operator (\ref{eq:302}) as is the case in Eq. (\ref{eq:301}), although in most situations in which one does not look at the $\gamma$-emission channel, this process will have little effect on the pair transfer cross section. Effects likely already taken into account in the imaginary part of the optical potential (op: reaction dielectric function) used to calculate the distorted waves. This is also the case regarding the 1$n$-channel Eq. (\ref{eq:1} (a)). 

\begin{figure}
	\centerline{\includegraphics*[width=8cm,angle=0]{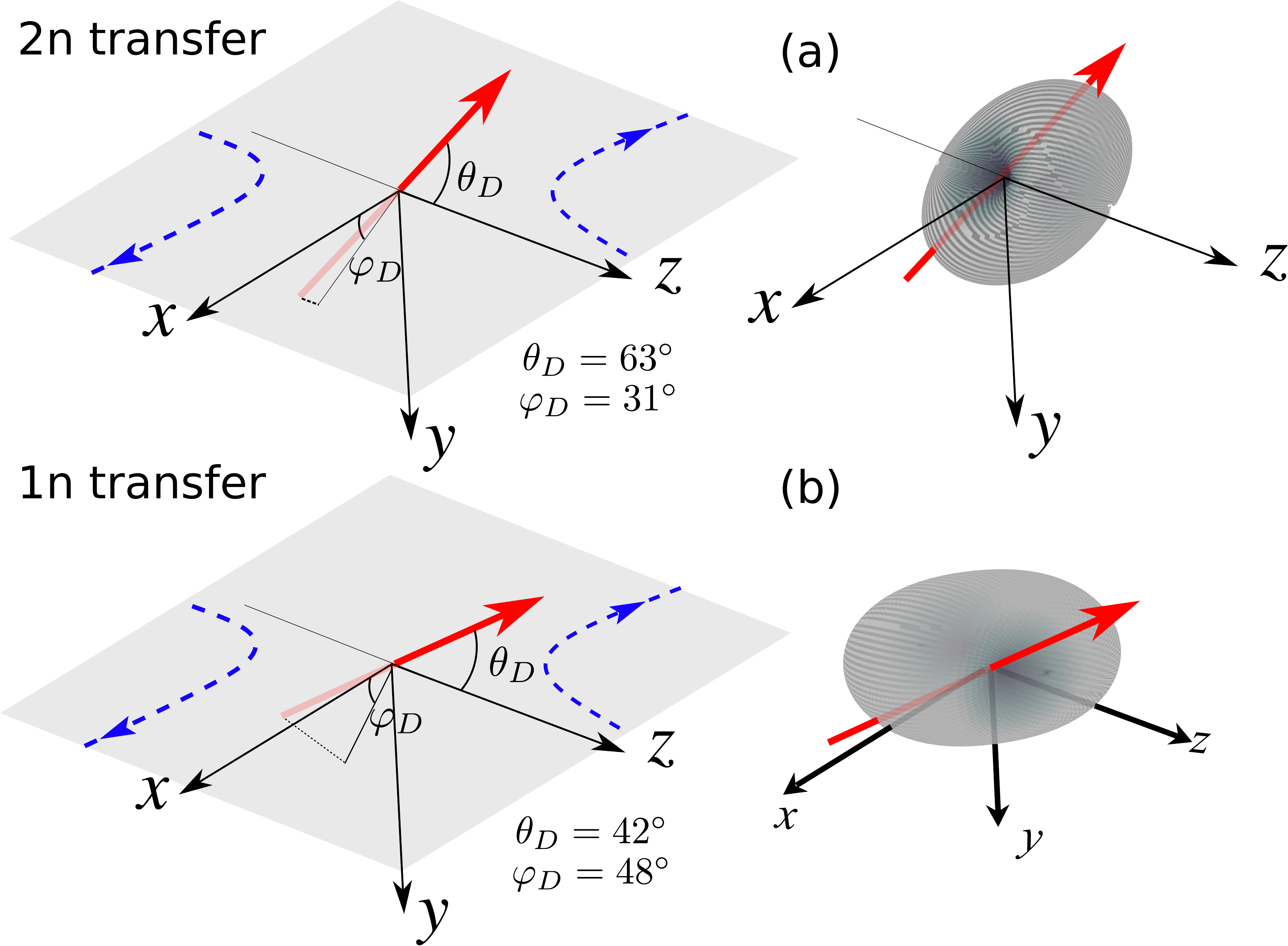}}
	\caption{({\bf{a}}) (left) representation of the dipole operator (\ref{eq:302}) associated with Cooper pair tunneling for $E_\gamma=4$ MeV, $E_{cm}=154.26$ MeV and $\theta_{cm}=140^\circ$. The dashed curves are schematic representations of the Sn (left corner) and Ni (upper right corner) trajectories, in the center of mass frame of reference; (right) angular radiation pattern in cartesian coordinates for the same $\gamma$-ray energy, and kinematic conditions; ({\bf{b}}) same as (a), but for one-neutron tunneling processes.}\label{fig2}
\end{figure}

If the photon polarization is not measured, the triple 2-nucleon transfer absolute  differential cross section can be written as
\begin{widetext}
\begin{align}
  \label{eq:303}
  \nonumber\frac{d^3\sigma_{2n}}{d\Omega_\gamma d\Omega dE_\gamma}&=\rho_f(E_f)\,\rho_\gamma(E_\gamma)\left(\left|\mathcal T^{1}(\mathbf k_\gamma,\mathbf k_f)\right|^2+\left|\mathcal T^{-1}(\mathbf k_\gamma,\mathbf k_f)\right|^2\right)\delta(E_i-E_\gamma-E_f+Q_{2n})\\
 &=\frac{\mu_i\mu_f}{(2\pi\hbar^2)^2}\frac{k_f}{k_i}\left(\frac{E_\gamma^2}{(\hbar c)^3}\right)\left(\left|\mathcal T^{1}(\mathbf k_\gamma,\mathbf k_f)\right|^2+\left|\mathcal T^{-1}(\mathbf k_\gamma,\mathbf k_f)\right|^2\right)\delta(E_i-E_\gamma-E_f+Q_{2n}),  
\end{align}
\end{widetext}
where $\rho_f(E_f)$ and $\rho_\gamma(E_\gamma)$ are the heavy ion and photon phase spaces respectively, while  $Q_{2n}$ is the reaction $Q$-value, and $E_i$, $E_f$ are the kinetic energies in the initial and final channels. 
The resulting ($2n$) radiation pattern is  displayed in Fig. \ref{fig2} (a)  in cartesian  coordinates, $z$-axis coinciding with the beam, the $x-z$ being the reaction plane.

The double differential cross section obtained by integrating Eq. (\ref{eq:303}) over all $\gamma$ angles,
\begin{align}
  \label{eq:306}
  \nonumber&\frac{d\sigma_{2n}}{d\Omega\,dE_\gamma}=\int \frac{d^3\sigma}{d\Omega_\gamma\, d\Omega\,dE_\gamma}\,d\Omega_\gamma=\frac{\mu_i\mu_f}{(2\pi\hbar^2)^2}\frac{k_f}{k_i}\left(\frac{8\pi}{3}\frac{E_\gamma^2}{(\hbar c)^3}\right)\\
         &\;\times\left(\left|T_1(\mathbf k_f)\right|^2+\left|T_{-1}(\mathbf k_f)\right|^2+\left|T_{0}(\mathbf k_f)\right|^2\right)\delta(E_i-E_\gamma-E_f+Q_{2n}),    
\end{align}
was used in the calculation of the ($2n$)  $\gamma$-strength function displayed in Fig. \ref{fig1} (a).

The strength of the two different photon polarizations  is provided by
\begin{align}
  \label{eq:2600}
  \left|\mathcal T^1(\mathbf k_\gamma,\mathbf k_f)\right|^2;\quad \left|\mathcal T^{-1}(\mathbf k_\gamma,\mathbf k_f)\right|^2, 
\end{align}
while the associated analyzing power can be written as
\begin{align}
  \label{eq:3600}
    \mathcal P(\mathbf k_\gamma,\mathbf k_f)=\frac{\left|\mathcal T^{1}(\mathbf k_\gamma,\mathbf k_f)\right|^2-\left|\mathcal T^{-1}(\mathbf k_\gamma,\mathbf k_f)\right|^2}{\left|\mathcal T^{1}(\mathbf k_\gamma,\mathbf k_f)\right|^2+\left|\mathcal T^{-1}(\mathbf k_\gamma,\mathbf k_f)\right|^2}.
\end{align}

\begin{figure}
  \centerline{\includegraphics*[width=8cm,angle=0]{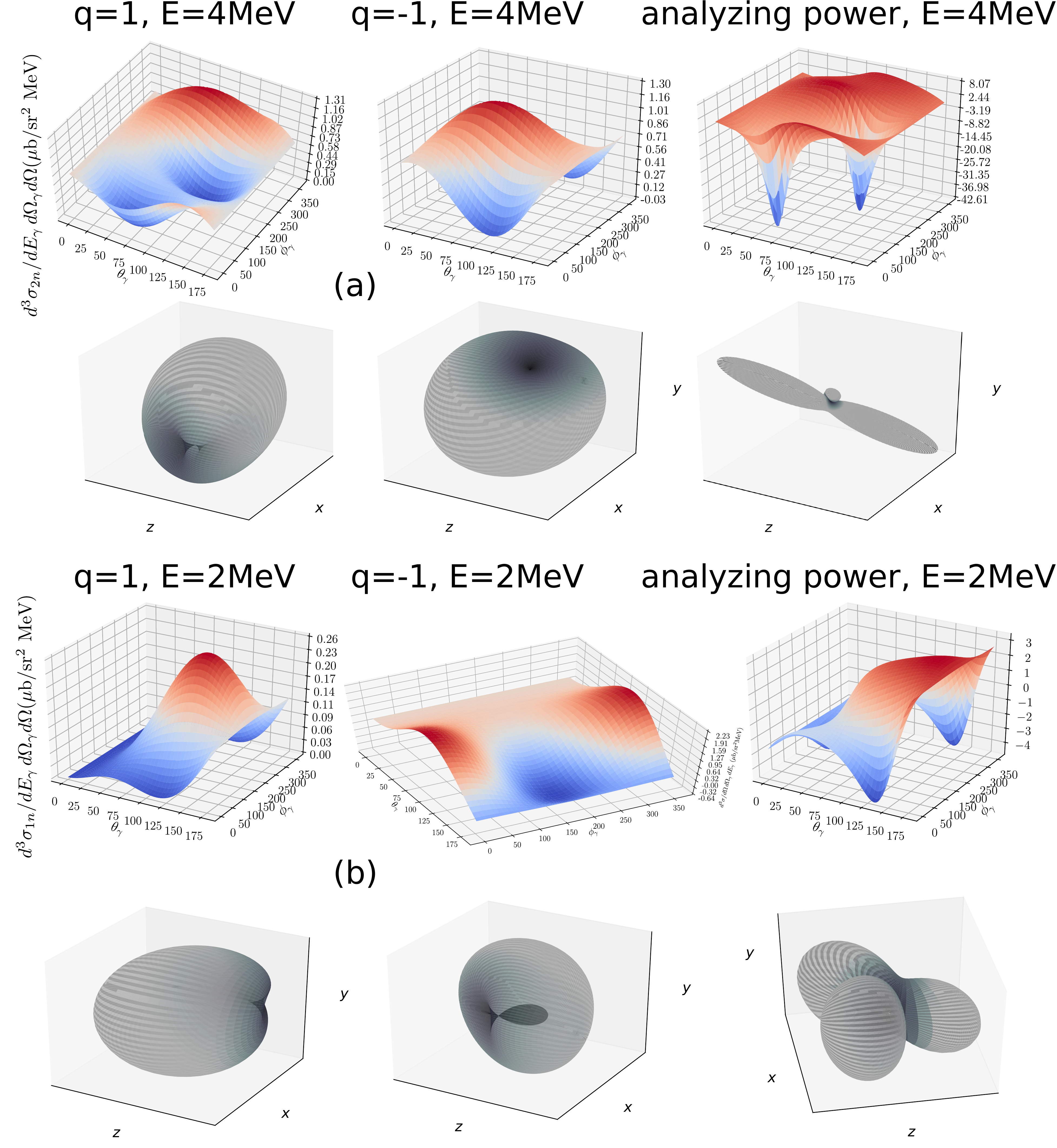}}
	\caption{({\bf{a}}) Polarization observables in both spherical and cartesian coordinates associated with the $\gamma$-emission of Cooper pair tunneling calculated with the same kinematic conditions as in the previous figures, and making use of Eqs. (\ref{eq:2600}) and (\ref{eq:3600}); ({\bf{b}}) Similar for one-neutron tunneling calculated making use of equations which parallel (\ref{eq:2600}) and (\ref{eq:3600}) and which take into account the coupling coefficients associated with the quasiparticles total angular momentum $j$ entering in the incoherent contributions to the $(1n)$-channel.}\label{fig3}
\end{figure}

The above quantities  for the $(2n)$-tunneling process (\ref{eq:1} (b)) are displayed in both spherical and cartesian coordinates in Fig. \ref{fig3} (a).

Making use of expressions similar to (\ref{eq:306}), (\ref{eq:303}) and (\ref{eq:306})-(\ref{eq:3600}), the $(1n)$ $\gamma$-strength function, angular distribution and polarization patterns were calculated, and are displayed in Figs.
\ref{fig1} (b), \ref{fig2} (b), and \ref{fig3} (b).

Although one is entering a new and unexplored field of research, two results look nonetheless evident: ({\bf{i}}) while the radiation pattern associated with the ($1n$)-transfer approaches that of a classical dipole (Fig. \ref{fig2} (b) right), the one corresponding to the $(2n)$-transfer reflects the subtle, quantum mechanical interweaving of the real and imaginary parts of the $T_{\pm1},T_0$ matrices appearing in the absolute, triple differential cross section (Fig. \ref{fig2} (a) right); ({\bf{ii}}) the difference between maxima and minima of the analyzing-power is short of an order of magnitude larger in the coherent $(2n)$-tunneling (Fig. \ref{fig3} (a)) than in the inclusive, incoherent ($1n$)-tunneling (Fig. \ref{fig3} (b)) and, as a result, the  extreme directional pattern  --fingerprint-- of coherent (Cooper pair)-transfer radiation. Both effects should be amenable to experimental test with present day magnetic and $\gamma$ spectrometers.

Concerning studies of polarized THz radiation from Josephson junctions in condensed matter physics, we refer to \cite{Ozyuzer:07} (see also \cite{Elarabi:17,Tsujimoto:20} and refs. therein) where peaks of polarized Josephson radiation have been  observed in a layered high $T_c$ superconductor (BSCCO), and unpolarized radiation at higher current and voltage bias were identified as thermal radiation. 
\section{Conclusions}
The incoherent summed contribution of quasiparticle states to the one-nucleon tunneling $\gamma$-strength function leads to a blackbody spectral radiance shape. The fact that the agreement is not perfect ($\chi^2=3.7\times10^{-3}$) tells us that, after all, the nucleus is not an infinite system. What is nonetheless striking is that already the incoherent contribution of essentially only four quasiparticle states populated within the range of $\approx 2$ MeV,  allows the system to display macroscopic, thermally equilibrated-like behaviour. At the antipodes regarding the physics at the basis of the $\gamma$-strength function associated with the tunneling of a Cooper pair between two $\ket{BCS}$ states made out of few equally phased Cooper pairs, which displays a Gaussian line shape with high accuracy ($\chi^2_G=1.8\times10^{-3}$).

Regarding the eventual measurements of $\gamma$-ray emission angular distributions and polarization (analyzing power), they are likely to become specific probes in the quest of gaining microscopic insight into the (ac) Josephson- and Joule-like effects within the framework of finite quantal many body systems of which the atomic nucleus constitutes a paradigm. Also to provide further evidence of the validity of BCS theory of superconductivity down to condensates of few Cooper pairs (3-6), and currents of a single Cooper pair.

From the vantage point of view provided by the major progress which has taken place in nuclear physics within the field of $\gamma$-detector arrays, so called $4\pi$-detectors, during the last three decades, it is not unrealistic to think that the predictions advanced in the present paper can undergo experimental test in the near future. We dedicate this paper  to the memory of Bent Herskind, pathfinder and master in this field of research. Following his example we allow ourselves  to speculate on what similar measurements to the ones advocated above within the field of nuclear physics, can provide of physical insight  regarding the expression of  quantum mechanics  at the macroscopic level, if carried out within the framework of condensed matter, fermionic BEC  physics.

{\it Acknowledgements}: R.A.B. acknowledges C. Pethick for inspiring discussions and suggestions.This work was performed under the auspices of the U.S. Department of Energy by Lawrence Livermore National Laboratory under Contract No. DE-AC52-07NA27344. This work is part of the I+D+i project with Ref. PID2020-114687GB-I00, funded by MCIN/AEI/10.13039/501100011033.

%	\bibliographystyle{unsrtnat}
	%\bibliography{C:/Gregory/book/nuclear_bib}
	%\bibliography{/home/gpotel/Desktop/Gregory/Articulos/Mios/libro_nuclear/nuclear_bib}
%	\bibliography{/home/gregory/book/nuclear_bib}
 \end{document}